\documentclass[a4paper]{article}

\pdfoutput=1

\usepackage[affil-it]{authblk}
\usepackage{graphicx,graphics}
\usepackage{combelow}

\usepackage{color}

\title{Ab initio investigation of optical properties in triangular graphene - boron nitride core-shell nanostructures} 
\author[1,2]{Andreea A. Nil\u{a}}
\author[1,3]{G. A. Nemnes}
\author[4]{A. Manolescu}
\affil[1]{University of Bucharest, Faculty of Physics, Materials and Devices for Electronics and Optoelectronics 
Research Center, 077125 Magurele-Ilfov, Romania}
\affil[2]{National Institute of Materials Physics,
077125 Magurele-Ilfov, Romania}
\affil[3]{Horia Hulubei National Institute for Physics and Nuclear Engineering,
077125 Magurele-Bucharest, Romania}
\affil[4]{School of Science and Engineering, Reykjavik University,
Menntavegur 1, IS-101, Reykjavik, Iceland}
\date{}

\begin{document}
\maketitle
\begin{abstract}
We calculate the optical properties of atomic-sized core-shell
graphene - boron nitride nanoflakes with triangular shaped crossection
using the density functional theory. The optical
properties can be tuned by using different sizes and proportions of the
core-shell materials. Anisotropic effects manifested in the absorption
of unpolarized light with different orientations of the optical vector
are pointed out.
\end{abstract}

\section{Introduction}

A lot of effort has been dedicated recently, both theoretically and experimentally, for the study of core-shell nanostructures. Examples of this kind are core-shell nanowires and nanoparticles \cite{ethayaraja,peng} which are the building blocks of numerous promising applications. Tuning the optical properties by controlling the band gap is a great challenge since
attention should also directed to avoid the interfacial defects and stress which may occur due lattice mismatches
between the two materials \cite{murphy}.

Graphene is a remarkable material that has enjoyed numerous studies
in the recent years, due to the special optical, electrical and mechanical
properties. One way to control the band gap is to resort to lower
dimensional systems or of finite size, such as quasi one-dimensional
nanoribbons \cite{han}, nanoflakes \cite{kuc}, or quantum dots \cite{Zhang2008,Guclu2014}.  
Another approach is to
create heterostructures with graphene on top of a boron nitride substrate.
Taking advantage of the similarity between the atomic structure of
graphene and of the hexagonal boron nitride (hBN), with a lattice mismatch
of only 2\%, hybrid graphene-hBN heterostructures have been synthetised
using chemical vapor deposition \cite{ci}.

In this paper we investigate the optical properties of core-shell graphene - hBN nanoflakes. Using density functional
theory calculations we determine the complex dielectric function and the other optical quantities of interest such as
the absorption coefficient and refraction index.
Our goal is to characterize the gap opening and the optical properties by adjusting the core-shell ratios.
In addition, we point out anisotropic properties of the system by modifying the orientation of the optical vector.

\vspace*{-0.3cm}
\section{Model and Method}

\begin{figure}[h]
  \centering
  \includegraphics[width=3.2cm, height=3.2cm]{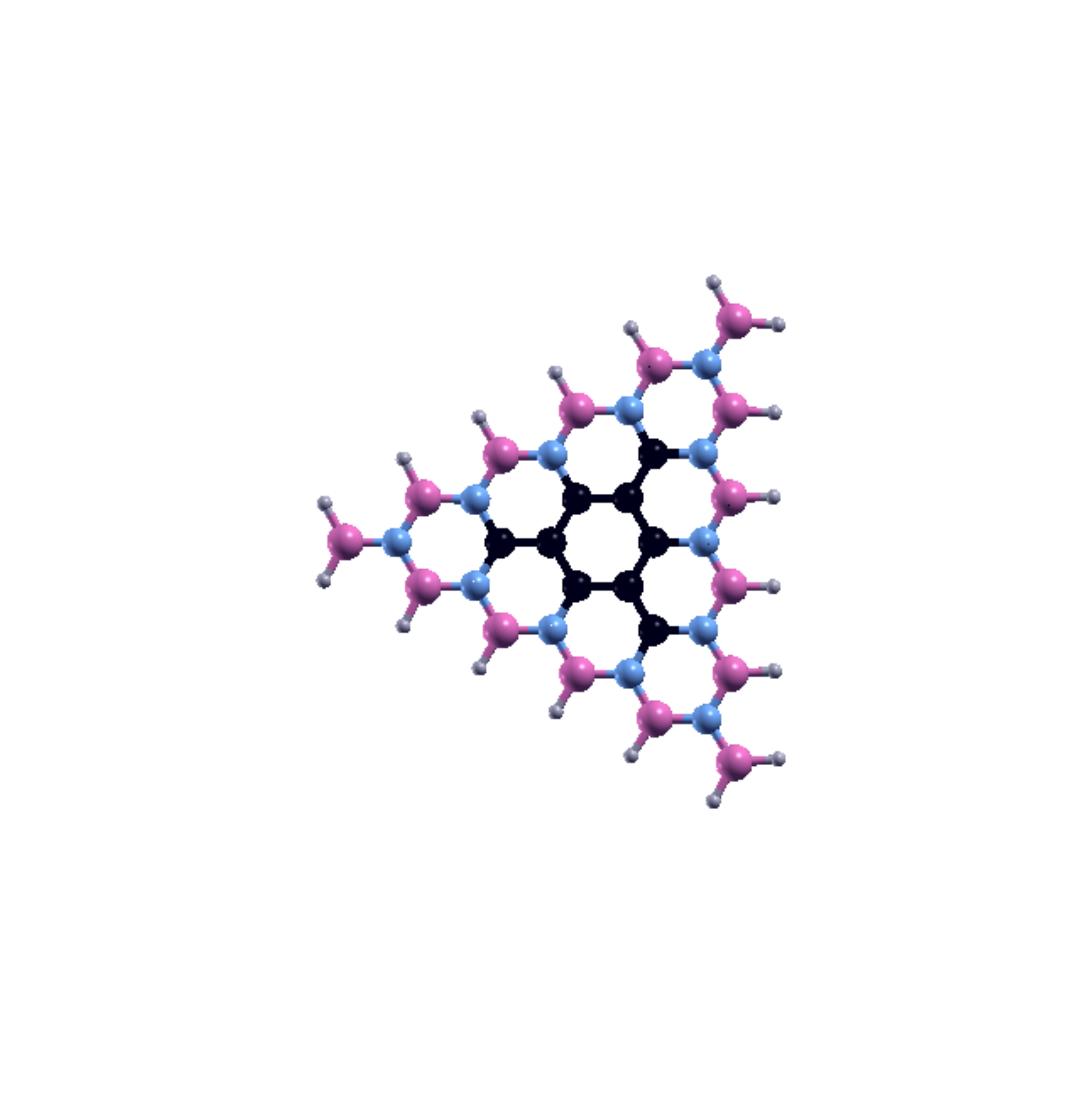}  \hspace{-0.5cm}              
  \includegraphics[width=3.2cm, height=3.2cm]{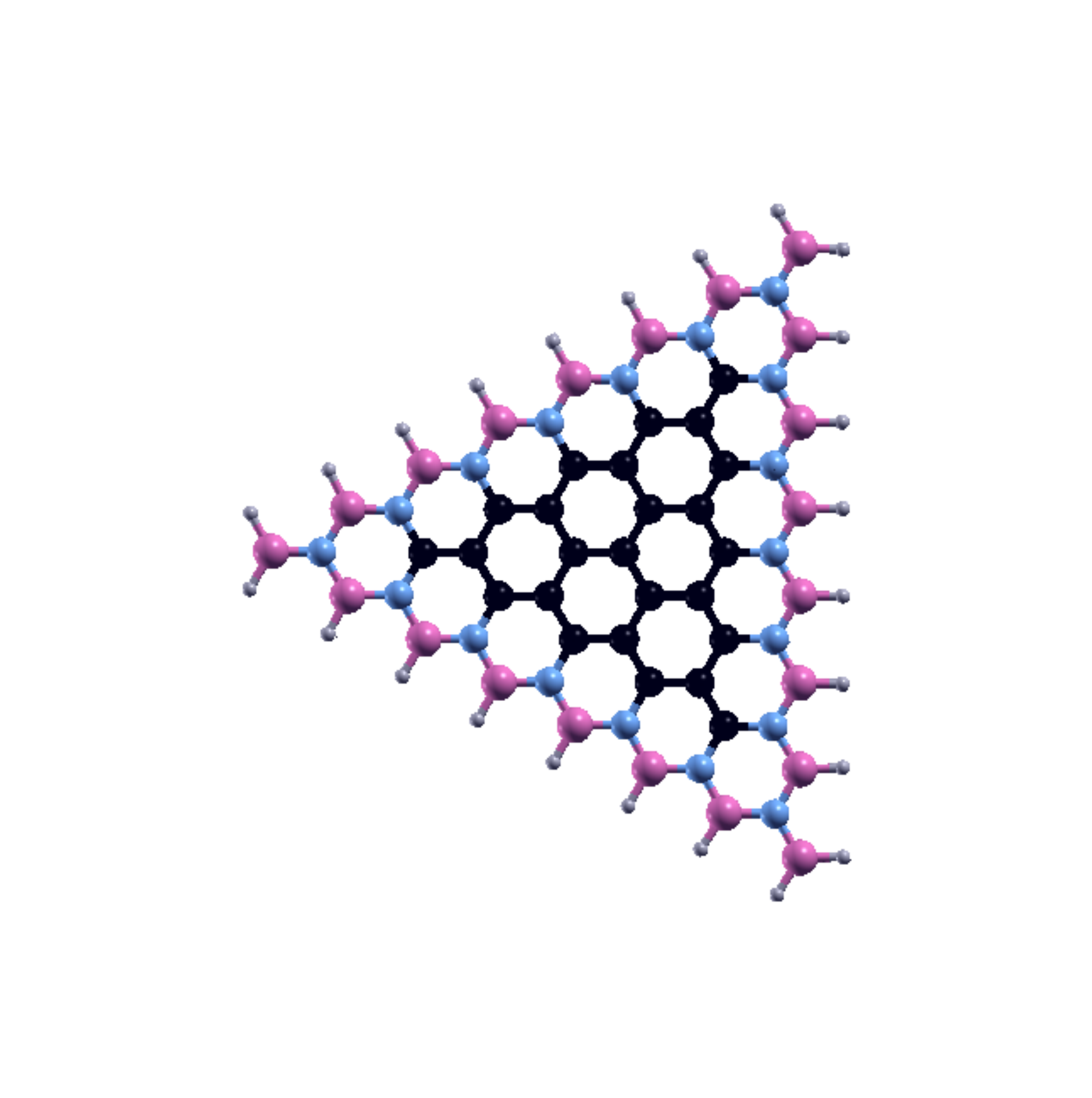}
  \includegraphics[width=3.2cm, height=3.2cm]{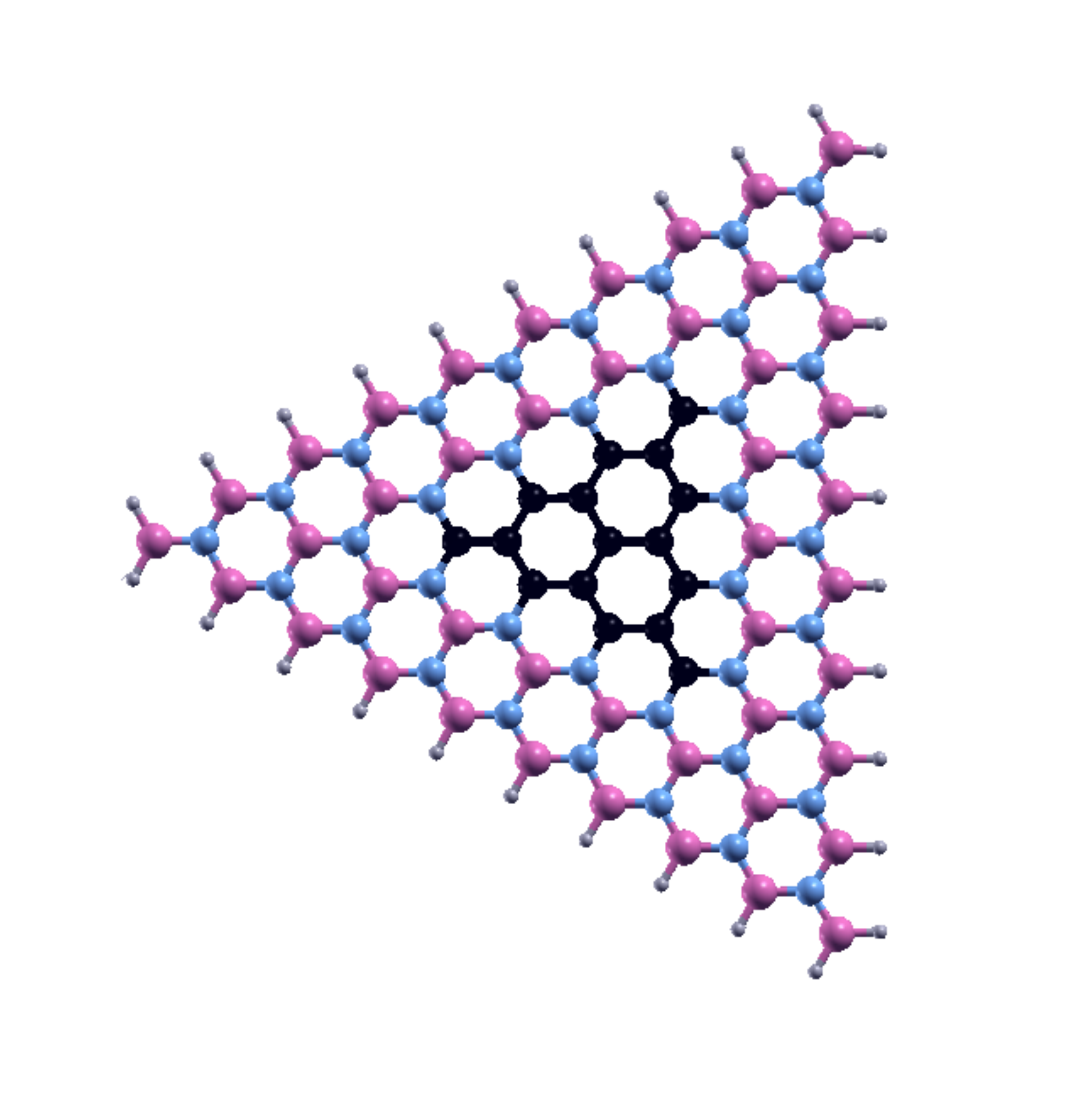}
  \includegraphics[width=3.2cm, height=3.2cm]{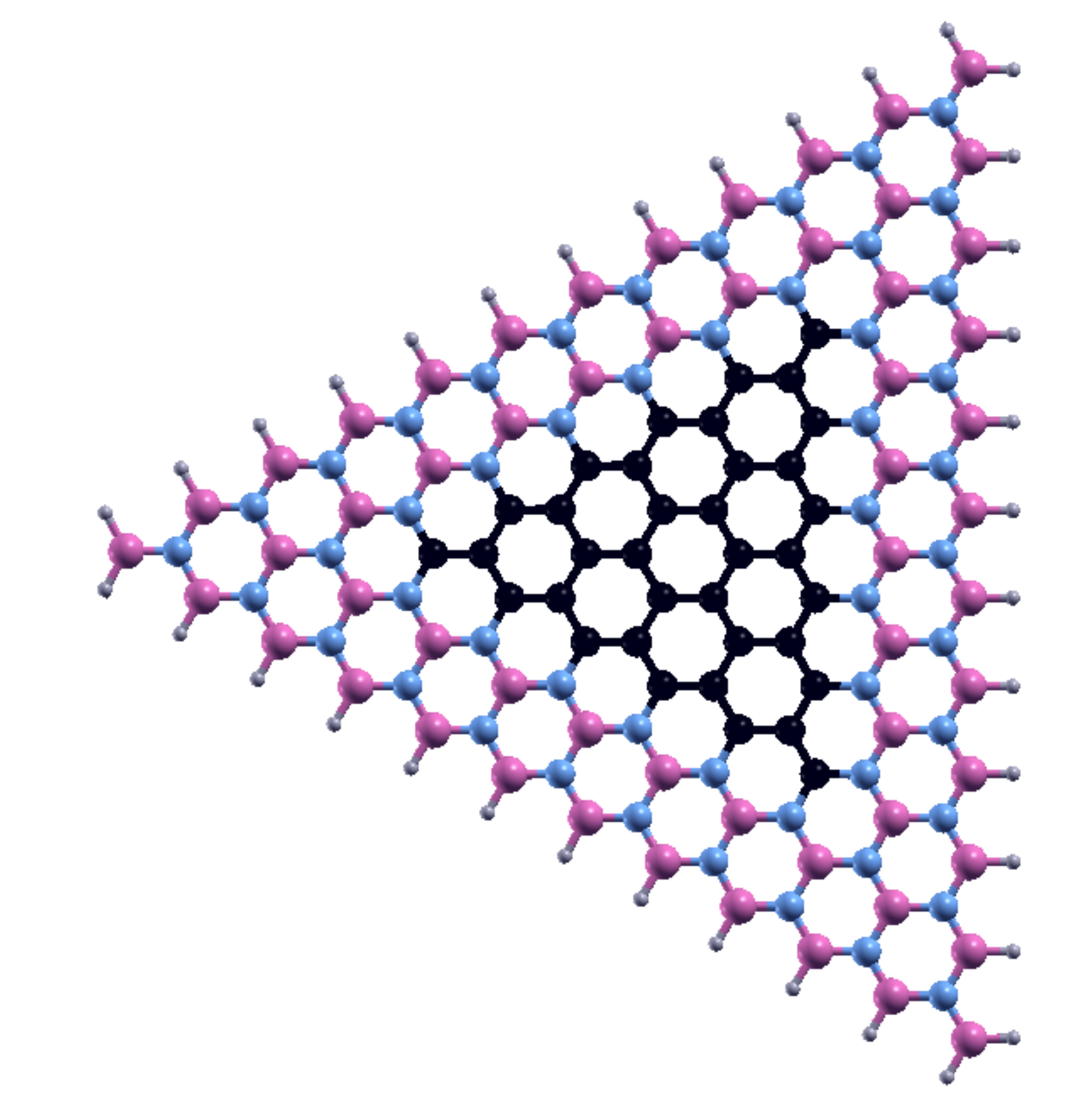}
  \caption{Graphene-hBN core-shell nanoflakes of different sizes: 54 (L4), 88 (L6), 130 (L8) and 180 (L10) atoms. The edges are hydrogen passivated. These nanoflakes are of G-BN type.}
 \label{pic1}
\end{figure}

We consider triangular graphene-hBN structures of different sizes as
depicted in Fig.\ \ref{pic1}. The graphene cores are surrounded by hBN
shell layers, which are hydrogen passivated. We label this class of
systems by G-BN. By exchanging the graphene with hBN we obtain similar,
but complementary structures, which we call BN-G.

The {\it ab initio} calculations are performed using the SIESTA
package \cite{11} and the approach is based on the Khon-Sham
self-consistent density functional theory in the local density
approximation (LDA).  The nanoflakes are placed in a cubic super-lattice
of linear size 40$a_0$, where $a_0=1.42$\AA \ is the distance between
two Carbon atoms in graphene, so that they do not interact for all
the considered sizes.  For the structural relaxations we considered
a double zeta polarized (DZP) basis set and a real space grid cutoff
of 100 Ry. The equilibrium configurations were obtained for
residual forces less than 0.04 eV/\AA.

The optical properties are calculated for unpolarized light for a given direction of the optical vector, 
chosen in-plane, in (100) and (010) directions, and perpendicular to the plane (001). The complex dielectric function
$\epsilon(\omega)$ is calculated and the other optical quantities are extracted: absorption coefficient $\alpha$,
reflectance, refractive index $\tilde n = n+i\kappa$ and optical conductivity $\sigma(\omega)$ using the
relation $\epsilon(\omega)=\epsilon_0+i\sigma(\omega)/\omega$, where $\epsilon_0$ is the vacuum permittivity.

\section{Results and Discussion}

%\subsection{Typical optical properties of G-BN systems}
%\vspace*{-0.3cm}

Despite being one atom thick the graphene monolayers absorb a rather large 2.3\% intensity of the visible light. The 
investigated core-shell nanoflake systems are a mixture of graphene, which is a highly conductive material, 
and hexagonal boron nitride, which is a wide band gap semiconductor. Therefore mixed optical properties are
expected.
\begin{figure}[h]
\centering
  \includegraphics[width=10.5cm, height=7.5cm]{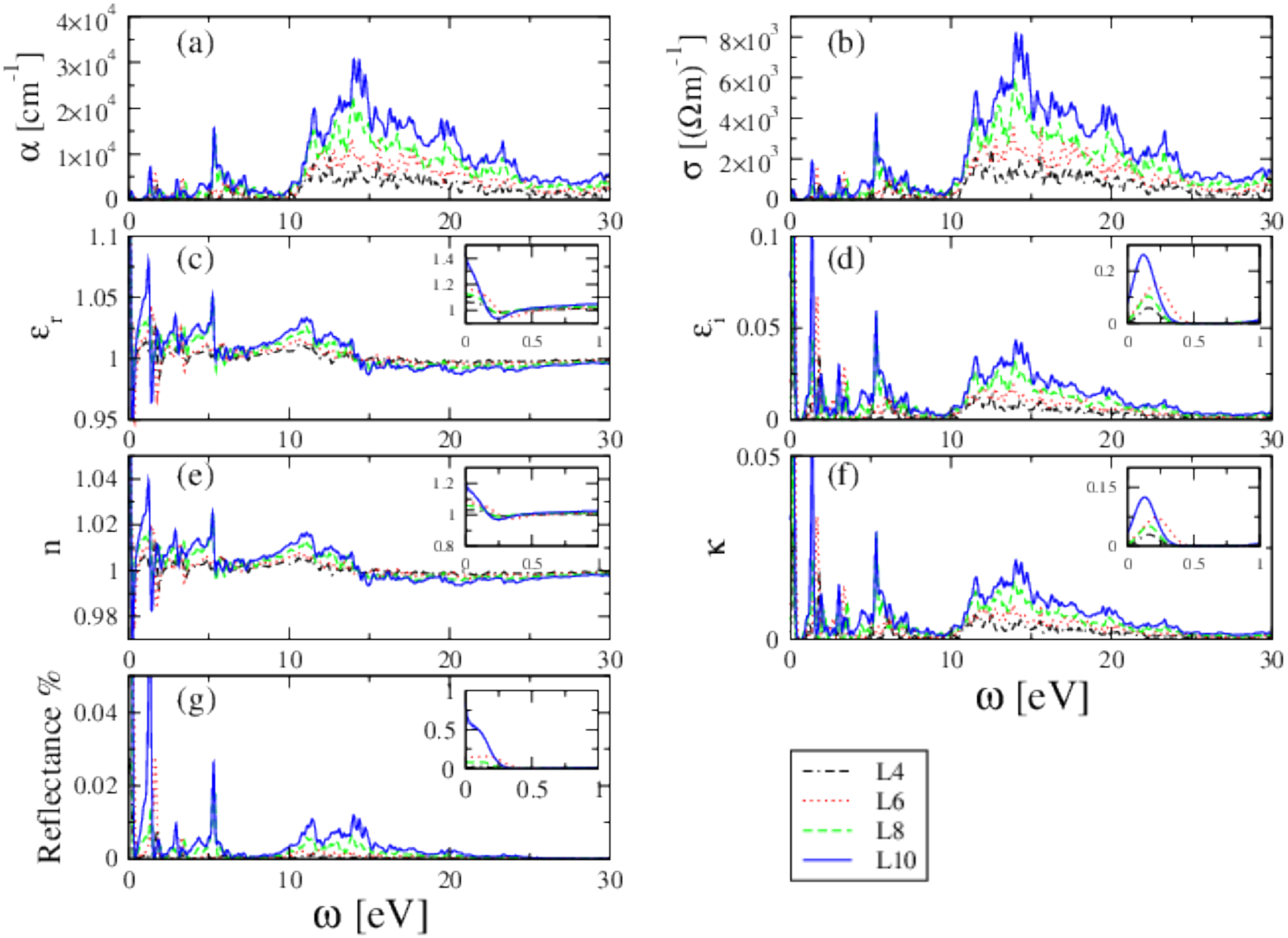} 
  \caption{Optical properties for different G-BN nanoflake sizes (L4-L10): (a) absorption coefficient; (b) optical conductivity; real part (c) and imaginary part (d) of the dielectric function; real part (e) and imaginary part (f) of the 
refractive index; (g) reflectance. The insets contain the data for a smaller energy range.
The size of each sample, L4-L10, is shown in Fig. \ref{pic1}.}  
  \label{pic2}             
\end{figure}
Figure\ \ref{pic2} shows optical spectra
of G-BN structures of different sizes for unpolarized light, with optical vector on the (100) direction.
Although the systems have rather small sizes, the optical properties are rather similar to those of 
the constituent bulk systems. Several features can be identified, which originate in the 
pristine infinite graphene and hBN systems. 
If the optical vector is perpendicular to the pristine graphene plane one observes a narrow peak at $\sim$4 eV,
which corresponds to the $\pi-\pi^{*}$ interband transition and a broader peak, with a prominent
absorption at $\sim$14 eV associated to the $\sigma-\sigma^{*}$.
For polarized light with the electric field perpendicular to the 2D plane, two peaks are observed 
in the high energy range at about 11 and 14 eV, corresponding to $\pi-\sigma^*$ and $\sigma-\pi^*$.
Using unpolarized light with optical vector (100) we obtain contributions from both sets of transitions
\cite{li,sedelnikova,13,14}.

%\subsection{Absorption coefficient for different core-shell ratios} 
%\vspace*{-0.3cm}

\begin{figure}[h]
\centering
  \includegraphics[height=6cm]{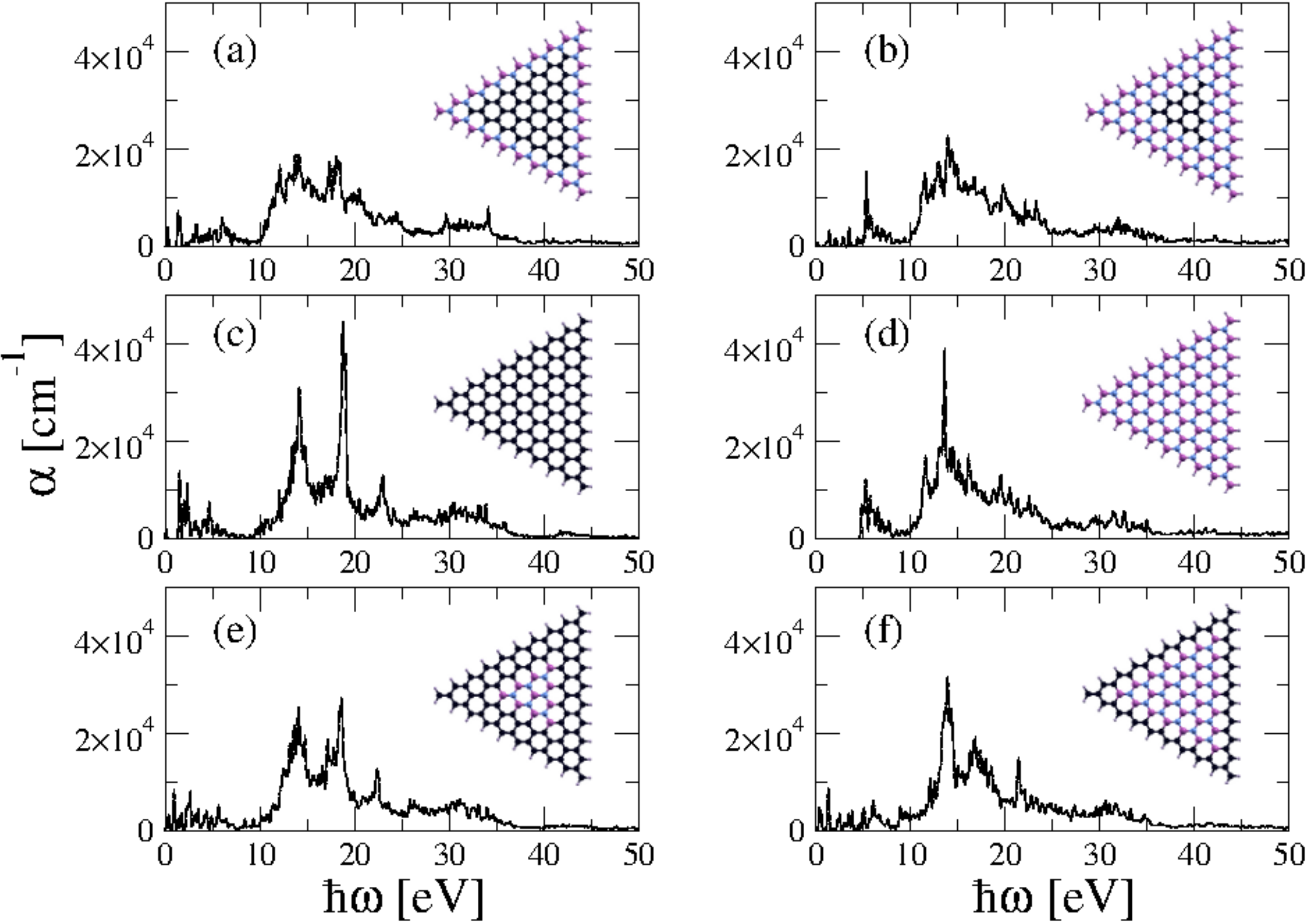} 
  \includegraphics[height=5.9cm]{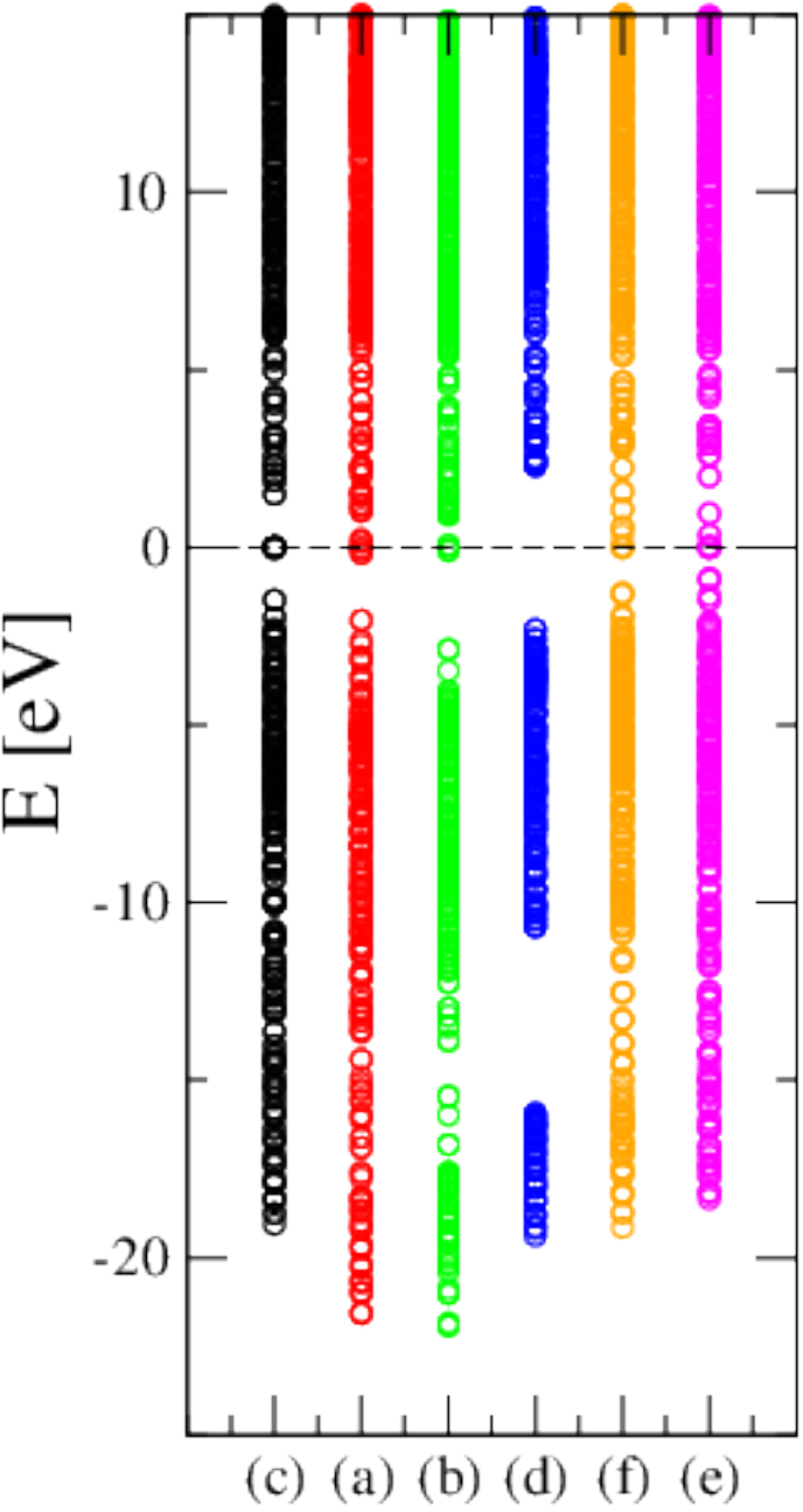} 
  \caption{Absorption coefficient $\alpha$ and eigenvalue spectra for different core-shell ratios.
The Fermi energy is marked by the horizontal dashed line.}  
  \label{pic3}             
\end{figure}

By changing the core-shell ratios, the band gaps of the systems can be tuned.  
In Fig.\ \ref{pic3} the absorption coefficient and the eigenvalue spectra are plotted for 6 structures, which gradually change from a pristine graphene nanoflake [Fig.\ \ref{pic3}(c)] to a hBN nanoflake [Fig.\ \ref{pic3}(d)].  
As before, unpolarized light is considered, with the 
optical vector in the (100) direction.
%\vspace*{-1cm}
%\vspace*{0.75cm}
For the pure hBN system a gap in the eigenvalue spectrum of $\sim$5 eV
is obtained. By introducing core or shell graphene layers, this gap
is reduced. In fact, in these systems the Fermi level crosses small groups of states
separated by mini-gaps, which are responsible for the weak absorption in the low frequency range.
The change is gradual and the features stemming from both materials are visible.

%\vspace*{-0.3cm}
%\subsection{Anisotropic effects}
%\vspace*{-0.1cm}
%
\begin{figure}[h]
\centering
  \includegraphics[height=5.0cm]{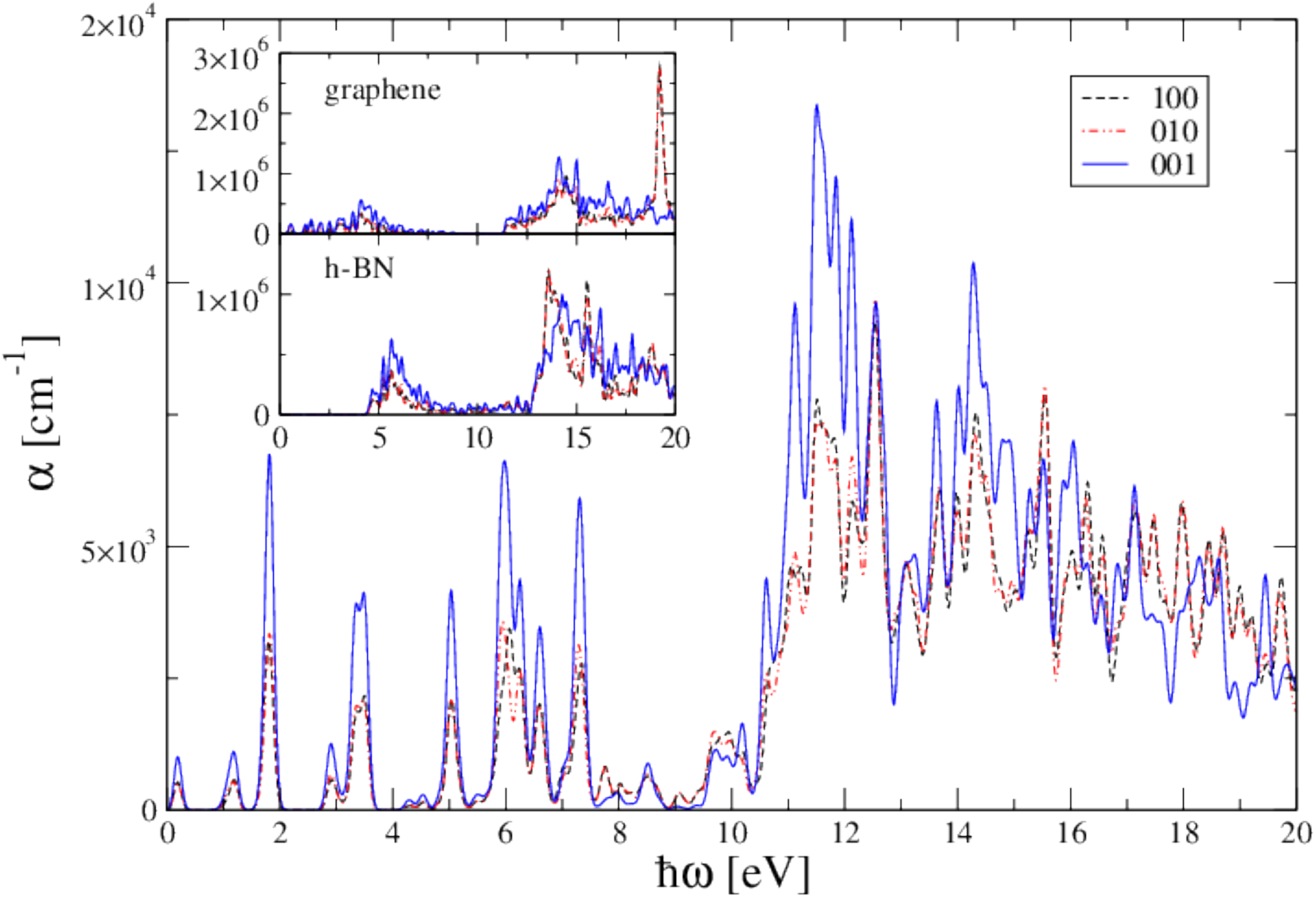} 
  \caption{Anisotropic effects introduced by unpolarized light with optical vectors with different orientations. The insets contain
similar data for the infinite layers of graphene and hBN.}  
  \label{pic5}             
\end{figure}
Next, we performed calculations on the G-BN L4 system by changing the orientation of the optical vector of unpolarized light, in-plane and perpendicular on the 2D surface. In Fig.\ \ref{pic5} the anisotropic effects can be pointed out
by changing the orientation of the optical vectors. 
A larger absorption coefficient is obtained for the (001) optical vector, while the in-plane orientations
 are mostly equivalent. Although the core-shell system is rather small, a connection with the absorption 
 spectra of graphene and hBN pristine systems can still be observed. 

\vspace*{-0.3cm}
\section{Conclusions}
\vspace*{-0.3cm}

In summary, we have carried out ab initio calculation of optical
properties for core-shell nanoflakes triangular cross-section in the
framework of density functional theory using the SIESTA software \cite{11}.
Our results indicate the possibility of changing 
the optical properties with increasing the system size or composition. Moreover,
the absorption coefficient can be changed by adjusting the proportions of the core-shell
materials of different band gaps, i. e. graphene and hBN. Since graphene displays
a zero band gap, by introducing BN either as core or shell material,
the nanoflakes present a gradual increase of the gap. 
Anisotropic effects are pointed out in the case of unpolarized light, by changing
the orientation of the optical vector for the core-shell system, in-plane
and perpendicular.\\

\vspace*{-0.3cm}
{\bf Acknowledgement}\\
Support of the EEA Financial Mechanism 2009-2014 staff mobility 
grant, the hospitality of Reykjavik University, and instructive discussions with
Sigurdur Ingi Erlingsson are gratefully
acknowledged. G. A. Nemnes also acknowledges support from ANCS under
grant PN-II-ID-PCE-2011-3-0960.

\end{document}